 \documentclass[preprint2]{aastex631}
\usepackage{xcolor}
\usepackage{amsmath}
\usepackage{graphicx}
\usepackage{hyperref}

\usepackage{parskip}
\shorttitle{High Contrast Imaging of BD+45$^{\arcdeg}$598}

\graphicspath{{./}{figures/}}

\begin{document}

\title{SCExAO/CHARIS Multi-Wavelength, High-Contrast Imaging of the BD+45$^{\arcdeg}$598 Debris Disk}

\author[0000-0001-5763-378X]{Maria Vincent}
\affiliation{Institute for Astronomy, University of Hawai'i at Mānoa, Honolulu, HI 96814, USA}

\author[0000-0002-6964-8732]{Kellen Lawson}
\affiliation{NASA Goddard Space Flight Center, Greenbelt, MD}

\author[0000-0002-7405-3119]{Thayne Currie}
\affiliation{Department of Physics and Astronomy, University of Texas at San Antonio, San Antonio, TX 78006, USA}
\affiliation{Subaru Telescope, National Astronomical Observatory of Japan, Hilo, HI 96720, USA}

\author[0000-0001-5058-695X]{Jonathan P. Williams}
\affiliation{Institute for Astronomy, University of Hawai'i at Mānoa, Honolulu, HI 96814, USA}

\author[0000-0002-1097-9908]{Olivier Guyon}
\affiliation{Subaru Telescope, National Astronomical Observatory of Japan, Hilo, HI 96720, USA}
\affiliation{Astrobiology Center, Osawa, Mitaka, Tokyo 181-8588, Japan}
\affiliation{Steward Observatory, The University of Arizona, Tucson, AZ 85721, USA}

\author[0000-0002-3047-1845]{Julien Lozi}
\affiliation{Subaru Telescope, National Astronomical Observatory of Japan, Hilo, HI 96720, USA}

\author[0000-0003-4514-7906]{Vincent Deo}
\affiliation{Subaru Telescope, National Astronomical Observatory of Japan, Hilo, HI 96720, USA}

\author[0000-0003-4018-2569]{Sébastien Vievard}
\affiliation{Subaru Telescope, National Astronomical Observatory of Japan, Hilo, HI 96720, USA}
\affiliation{Astrobiology Center, Osawa, Mitaka, Tokyo 181-8588, Japan}

\begin{abstract}
We present a multi-wavelength (1.16$\mu$m-- 2.37$\mu$m) view of the debris disk around BD+45$^{\arcdeg}$598, using the Subaru Coronagraphic Extreme Adaptive Optics system paired with the Coronagraphic High Angular Resolution Imaging Spectrograph. With an assumed age of 23 Myr, this source allows us to study the early evolution of debris disks and search for forming planets. We fit a scattered light model to our disk using a differential evolution algorithm, and constrain its geometry. We find the disk to have a peak density radius of $R_0 = 109.6$ au, an inclination of $i = 88.1\arcdeg$, and position angle $PA = 111.0\arcdeg$. While we do not detect a substellar companion in the disk, our calculated contrast limits indicate sensitivity to planets as small as $\sim 10 M_{\rm Jup}$ at a projected separation of 12 au of the star, and as small as $\sim 4 M_{\rm Jup}$ beyond 38 au. When measuring intensity as a function of wavelength, the disk color constrains the minimum dust grain size within a range of $\sim0.13$ to 1.01 $\micron$.

\end{abstract}

\section{Introduction}

As a pre-main sequence star condenses out of a large cloud of gas and dust through gravitational collapse, additional material remains as a flattened gas-rich ``protoplanetary'' disk to conserve angular momentum \citep{williams2011}. First through electrostatic interactions, followed by collective self-gravity, the dust grows to larger and larger sizes, ultimately becoming icy and rocky planetesimals. Over several million years, as the primordial disk is used up, planetesimal collisions re-populate the disk with a second generation of dust. 
\citep{wyatt2008,matthews2014,hughes2018}.
These gas-depleted, dust-rich debris disks can persist for tens of millions of years up to a billion years as they are continually enriched by fragmentation of bodies, often referred to as a ``collisional cascade''.
\citep{kenyon2004,kenyon2008,pan2012,wyatt2005,mustill2009}.
They are observable through the scattering of starlight at short wavelengths and thermal emission at longer wavelengths, and their study informs us of how long it takes planets to form, where they are, and what they are made of.

Observations of debris disks at different wavelengths highlight different aspects of the constituents and structure of the disk. Because of its large emitting area, the dust is easier to see around young stars than the planets themselves, and the disks' structural features (which may arise from a planet's interaction with a disk) and location around the star can help us constrain the presence of planets \citep{gibbs2019}. At millimeter wavelengths, the disk outshines the host star and we can observe the larger dust grains (so-called ``pebbles'') that fall down to the midplane. Although the star is brighter than the disk in the near-infrared (NIR), the short wavelengths reveal the smaller micron-sized dust grains that are swept along with the gas. This provides a complementary view of the disk structure that is highly sensitive to disturbances caused by disk instabilities, tidal forces from neighboring stars, and exoplanets. 

 NIR high contrast imaging suites has facilitated direct observations of disks by a confluence of instrumentation and post-processing techniques. Coronagraphy is used to reduce the starlight directly and differential imaging isolates the physical and instrumental effects. Differential imaging techniques like angular differential imaging \citep{marois2006} and spectral differential imaging \citep{sparks2002} are employed to extract the light of the bright central source from the circumstellar disk by modeling and subtracting the starlight in a data sequence. However, these methods are imperfect and the retrieved light is often attenuated due to over-subtraction of the disk signal. This attenuation can be quantified using forward modeling of synthetic scattered light models of the disk. These models help constrain the geometry of the disk and analyze the properties and composition of the dust. 

In this paper we look at BD +45$\arcdeg$598, a young Li-rich K1 type star located at 73.18 $\pm$ 0.12 pc \citep{gaia2020}, with a luminosity of 1.4$L_\odot$ and temperature $5280 \pm 100$K first described by \cite{guillout2009} in a spectroscopic survey of ROSAT All-Sky Survey (RASS) X-ray sources. The debris disk around it was discovered through high contrast imaging in 2.2$\mu$m using NIRC2 at the Keck II telescope by \cite{hinkley2021}. It is a nearly edge-on disk with an inclination of $\sim$87$^\circ$. As a member of the $\beta$ Pictoris moving group with an age of $23\pm 3$\,Myr, this source is well suited to study the early evolution of debris disks and search for forming planets. Its young age and the K1 spectral type make it a suitable target to investigate the formation of a planetary system about a late-type star analogous to the solar system \citep{hinkley2021}. Here, we observe this disk using the high contrast imaging capabilities of Subaru Coronagraphic Extreme Adaptive Optics \citep[SCExAO;][]{jovanovic2015} on Subaru and angular differential imaging. The light is fed through an integral field unit, Coronagraphic High Angular Resolution Imaging Spectrograph \citep[CHARIS;][]{groff2016}, that enables us to image the disk in multiple wavelengths. Through forward modeling, we determine the geometry of the disk, resolve color gradients across it, and search for planetary mass companions within the CHARIS field of view.

\section{Data}

\subsection{Observations}


BD +45$\arcdeg$598 was observed with Subaru on October 18 2021 using the SCExAO system paired with the CHARIS integral field spectrograph operating in low resolution (R$\sim$20) broadband (1.16- 2.37 $\mu$m) mode with a pixel scale of 0\farcs{}01615 pixel$^{-1}$ \citep{currie2022a}. The data were taken under good seeing conditions of 0\farcs4 and a wind speed of about 8.7 m/s. The dataset includes 52 exposures-- 49 science frames and 3 sky frames for sky subtraction. The exposure time for the first frame was 45.7 s and for the remaining frames, 60.5 s each, and thus the total integration time was a little over 52 minutes. Over the course of this observation, there was a 27.6$\arcdeg$ parallactic angle rotation for angular differential imaging (ADI).

\subsection{Data Reduction}


The raw data were passed through the CHARIS Data Reduction Pipeline \citep{brandt2017} to produce images with a size of $201\times 201$ pixels  ($2\farcs07\times2\farcs07$) for each of the 22 broadband wavelength channels. These data cubes were then further reduced using the CHARIS Data Processing Pipeline \citep{currie2011,currie2018}. The processing included sky subtraction, image cube registration, and spectrophotometric calibration (using the Kurucz stellar library) prior to point spread function (PSF) subtraction. The stellar flux used for the data reduction is normalized to the observed 2MASS H-band magnitude. Once the sky was subtracted from the extracted data, each slice of every data cube was recentered.

The PSF of the star was then removed to reveal the fainter disk signal. PSF subtraction was performed using ADI with the Adaptive, Locally Optimized Combination of Images (A-LOCI) algorithm. A-LOCI is used to construct an optimised reference PSF image as a linear combination of PSFs from a reference library \citep{currie2012,lafreniere2007}. The image of the disk is divided into smaller subsections from which the PSF image optimised to each of them is subtracted. The coefficients of the linear combination of reference images is calculated for each subsection, or optimization zone, to minimise speckle noise, and to ensure that any additional feature, for instance a substellar companion, doesn't alter the subtraction in other parts of the disk \citep{lafreniere2007,soummer2011,currie2012}.

In the A-LOCI procedure, we considered a range of values for the algorithm parameters to optimize the subtraction process and maximize the disk signal. We varied:
\begin{enumerate}
    \item Optimization area ($N_A$): This is the area of the image over which we try to solve for the coefficients characterized in terms of radial length and azimuthal width.  Measured in PSF cores, it is kept small enough that it represents the noise over the subtraction region, and large enough that the algorithm does not remove the signal from any detection more than the speckle noise.
    \item Rotation gap ($\delta_{\rm FWHM}$): Also measured in PSF cores, it excludes frames with small parallactic angle motion  from the reference image construction, and is kept large enough to limit self-subtraction.
    \item Geometry of subtraction \& optimization zones ($geom$): $geom$ is defined by the ratio of the radial and average azimuthal width of each of the optimization zones. Since this is an edge-on disk with good field rotation, we choose a value less than 1.
    \item Singular Value Decomposition (SVD) cutoff (SVD$_{\rm lim}$): It defines the limit below which the diagonal of the inverted covariance matrix will be set to 0.
\end{enumerate} 

The other parameters that were kept fixed were:
\begin{enumerate}
    \item $n_{\rm ref}$: Number of reference frames from which we construct the reference PSF, typically kept equal to the number of science frames.
    \item $r_{\rm min}$, $r_{\rm max}$: Minimum and maximum radius of subtraction measured in pixels 
    \item $\Delta r$: Defines the width of each subtraction section in the radial direction. The shape of the subtraction section is defined by $geom$.
\end{enumerate}

\begin{figure*}[h]
    \centering
    \includegraphics[trim=1cm 0.8cm 2.2cm 1.5cm, clip, width = \textwidth]{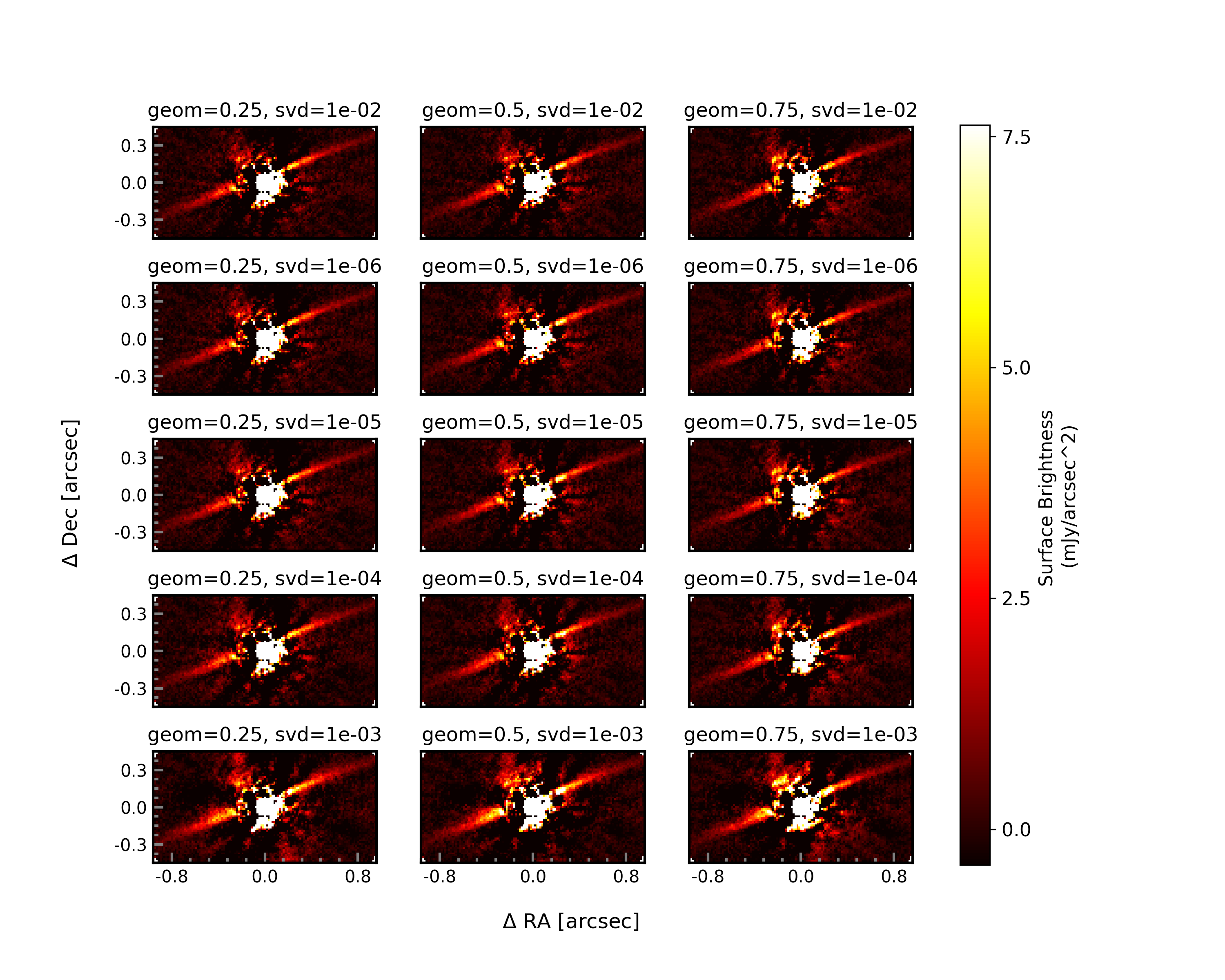}
     \caption{A grid of wavelength collapsed images of the disk following ADI A-LOCI PSF subtraction for different pairs of $geom$ and SVD$_{\rm lim}$ (indicated here as svd) values. $geom = 0.25$, SVD$_{\rm lim} = 10^{-5}$ has the highest signal to noise ratio along the spine of the disk.}
    \label{fig:gridI}
\end{figure*}

\begin{figure*}
    \centering
   \includegraphics[trim=1.7cm 0.2cm 3.7cm 0.2cm, clip, width = \textwidth]{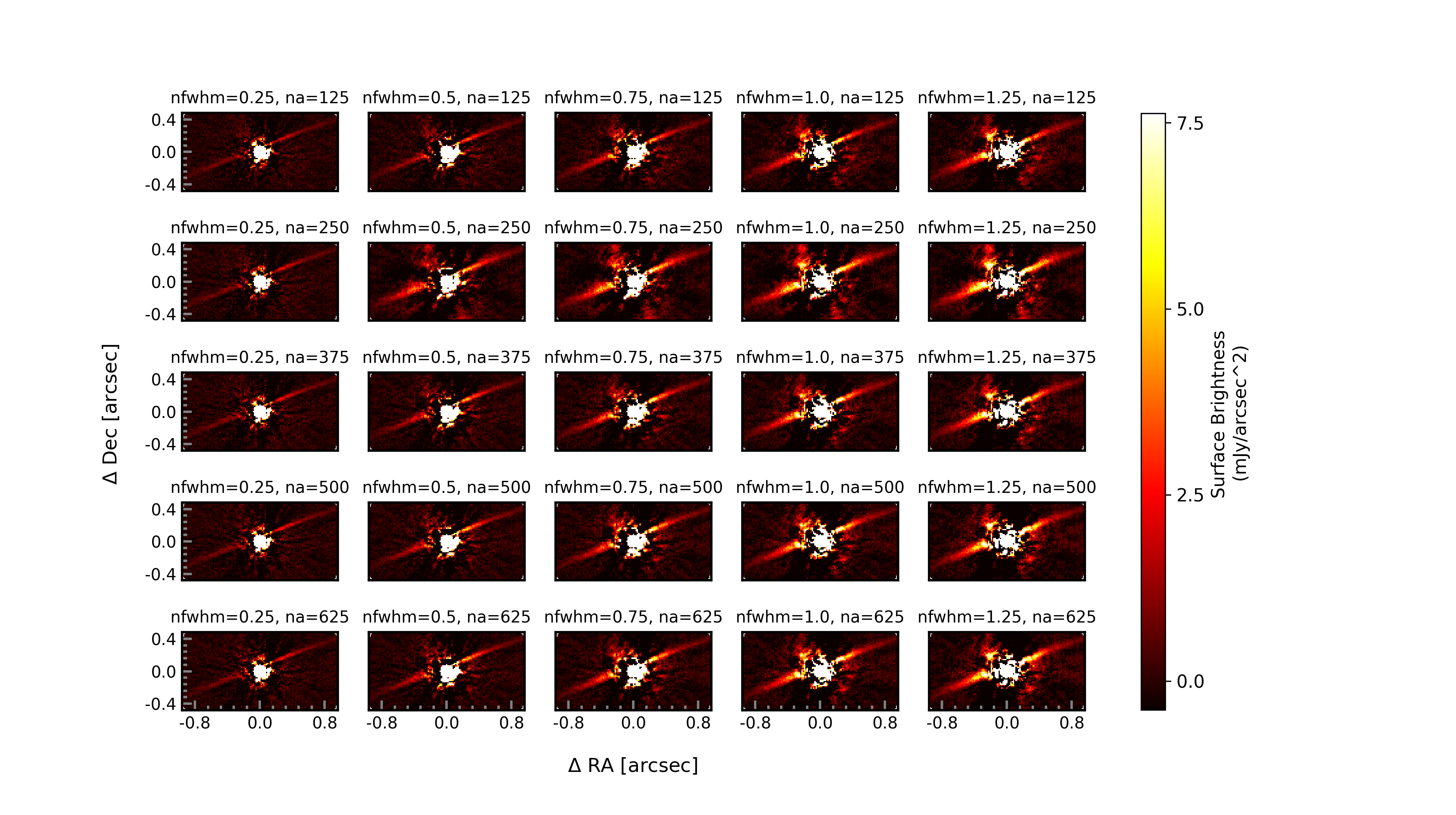}
    \caption{A grid of wavelength collapsed images of the disk following ADI A-LOCI PSF subtraction for different pairs of $N_A$ and $\delta_{\rm FWHM}$ (indicated here as nfwhm) values. $N_A = 250$, $\delta_{\rm FWHM}=0.75$ has the highest signal to noise ratio along the spine of the disk.}
    \label{fig:gridII}
\end{figure*}

We chose parameter values that resulted in PSF subtracted images with the highest signal to noise ratio (SNR) over the spine of the disk. For this, 2 grids of wavelength collapsed images (Figures \ref{fig:gridI} and \ref{fig:gridII}) and their corresponding SNR maps were created for different values of some pairs of parameters and the pixel count at various points on the disk was inspected across the images. The optimal algorithm settings for these parameters are as follows: $N_A = 250$, $\delta_{\rm FWHM}=0.75$, $geom = 0.25$, SVD$_{\rm lim} = 10^{-5}$, $n_{\rm ref} = 49$, $r_{\rm min} = 5$, $r_{\rm max} = 65$, and $\Delta r = 5$.

\subsection{Results from Data Reduction}

\begin{figure*}
    \centering
 \includegraphics[trim=3cm 2.5cm 5.4cm 2.5cm, clip, width=\textwidth]{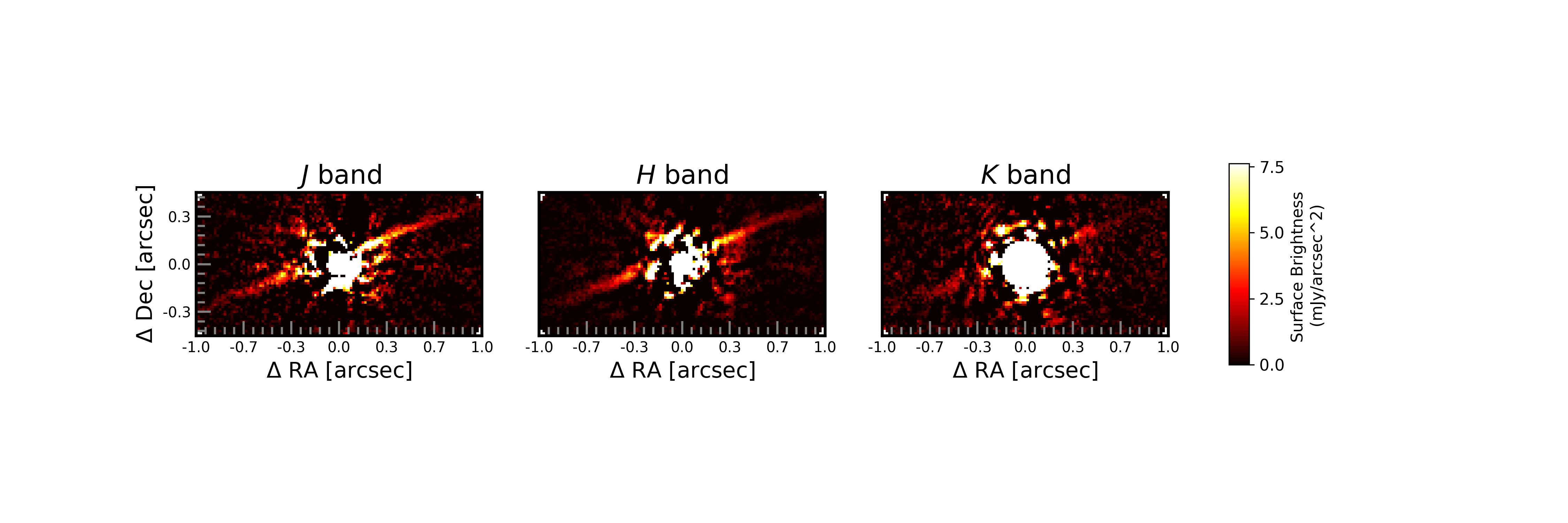}
      \caption{Images of the disk post ADI-ALOCI PSF subtraction with wavelength channels combined for images within J, H, and K bands}
    \label{fig:jhk_imgs}
\end{figure*}

\begin{figure*}
    \centering
      \includegraphics[trim=0.5cm 2cm 0cm 3cm,clip,width=0.8\textwidth]{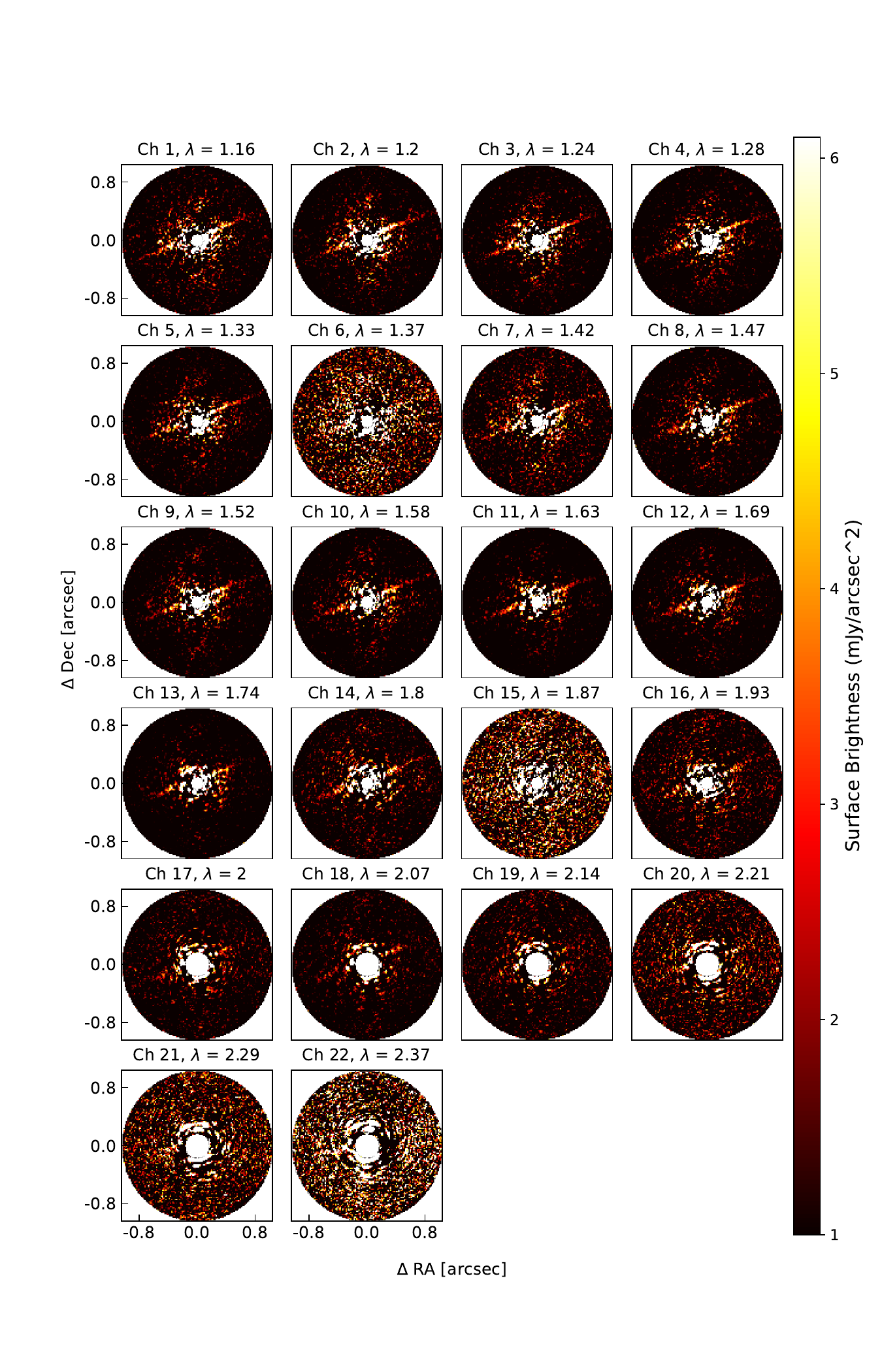}
      \caption{Images of the disk post ADI-ALOCI PSF subtraction for each wavelength channel. The wavelength of each channel is in microns.}
    \label{fig:reduc_channels}
\end{figure*}

Figure \ref{fig:jhk_imgs} shows the median-combined PSF subtracted images in the J (channels 1-6; 1.16-1.37 $\mu$m), H (channels 9-14; 1.52-1.8 $\mu$m), and K (channels 17 and 18; 2.00 and 2.07 $\mu$m) bands. The excluded channels are affected strongly by telluric absorption and lie outside the nominal J, H, and K bandwidths.
The ADI-ALOCI subtraction technique shows the detection of the disk across most wavelength channels (Figure \ref{fig:reduc_channels}). While the reduction process is not perfect and the disk cannot be seen in all channels, the wavelength collapsed images across the J, H, and K bands show the disk fairly well. The SNR varies from $\sim$2.8-7.5 along the disk (as measured in the wavelength collapsed image). The detection is the strongest in the H band and weakest in the K band. One reason can be that since intensity of scattered light is inversely proportional to fourth power of wavelength (Rayleigh Scattering), and the disk is visible due to the light scattered off the dust grains, the detection can be expected to be weaker at longer wavelengths. 


\section{Modeling the Disk}

\subsection{Disk Forward modeling} \label{sec:fwd_mod}

During PSF subtraction, there is some unavoidable loss of disk signal that biases our result. This needs to be taken into account while determining what true disk parameters may produce the diminished disk we see in our reduced data. Hence, in order to understand the morphology of this debris disk, we use forward modeling of a synthetic disk image. We create a model of the scattered light from a disk using the GRaTeR software \citep{augereau1999}, convolve it with the instrumental PSF for each of the 22 wavelength channels and then induce the signal loss that happens during the A-LOCI PSF subtraction process due to over-subtraction (where the presence of disk signal results in an overly bright PSF model) and self-subtraction (when the disk flux is captured in the reference frame results in a decrease of disk flux in the residual image after the reference subtraction) of the imaged disk signal as outlined in \cite{currie2017} and \cite{currie2019}. The instrumental PSF is empirically determined by taking the median combination of normalized cutouts of satellite spots at the 22 wavelengths throughout the entire data sequence \citep{lawson2021}. We then take the square of the difference of the model image and the wavelength collapsed, PSF-subtracted image of the disk, weighted by the noise map, to assess the goodness of fit. After binning this map to the size of the instrumental PSF, it is used to compute the reduced $\chi^2$, or $\chi_\nu^2$, metric. $\chi_\nu^2$ is calculated within a rectangular region \citep{lawson2020} centered on the masked out star of dimensions $125 \times 50$ px and rotated $-20\arcdeg$. 

Debris disks are seen due to photons hitting dust grains. Light hitting dust particles can be absorbed and re-emitted or scattered unevenly in all directions. This is based not only on the properties of the incoming photon, but also those of the dust grain, such as its size, shape and composition.
To model the scattering, we use a modification of the Henyey-Greenstein scattering phase function \citep[SPF;][]{henyeygreenstein1941}. This is conventionally defined by a single asymmetry parameter $g$, but has been shown to fail at reproducing scattered light images at small angles \citep{hughes2018}. We therefore use a linear combination of HG-SPFs based on observations of zodiacal dust \citep{hong1985}. The Hong model has three asymmetry parameters, $g_{\rm 1} = 0.7$, $g_{\rm 2} = -0.2$, and $g_{\rm 3} = -0.81$, with weights $w_{\rm 1} = 0.665$, $w_{\rm 2} = 0.330$, and $w_{\rm 3} = 0.005$.

We model the disk as a ring described by four free parameters which were explored using the differential evolution (DE) algorithm \citep{storn1997} following the procedure outlined in \citet{lawson2020}. DE is a genetic optimization algorithm that works by iteratively improving a population of trial solutions over successive ``generations''. For each generation, DE attempts to replace each population member with a trial solution, formed by adding the weighted difference of two random population members to the current best solution. Any trial solution that results in an improvement in the fitness metric (i.e. reduces $\chi_\nu^2$) replaces the corresponding member of the population. DE has been shown to converge quickly and is a more heuristic and robust method for optimising values within a parameter space \citep{storn1997,lampinen2004}. This finds an optimal solution faster than traditional grid search and is more computationally efficient than a Markov Chain Monte Carlo (MCMC) search.
 
The metaparameters for DE that govern how the algorithm works are as follows:
\begin{enumerate}
    \item $N_{pop}$, the number of population members varying for each parameter in one DE generation. The number of models created in one generation would be $N_{pop}$ times number of free parameters which, in this case, is $15 \times 4 = 60$
    \item $G_{min}$, the minimum number of generations before population convergence is allowed, which typically is 10.
    \item $\delta_{\sigma X}$ and $\delta_{\sigma FX}$, standard deviation thresholds that define the criteria for convergence. For the population to converge, $\delta_{\sigma X}$ is the maximum allowed standard deviation among normalised parameter values. $\delta_{\sigma FX}$ is the maximum possible value for the standard deviation of the fitness of the current population, divided by the median fitness of the current population \citep{lawson2021}. Both are set to 0.05 here.
\end{enumerate}

The disk model is described by four free parameters and few fixed parameters. The free parameters optimized through DE are:
\begin{enumerate}
    \item $R_0$, the radius of peak grain density in au
    \item $H_0/R_0$, the ratio of the disk's scale height at $R_0$ to $R_0$
    \item $i$, the disk inclination in degrees
    \item $PA$, the position angle of the disk in degrees.
\end{enumerate}

The following parameters are chosen to be fixed:
\begin{enumerate}
    \item $\alpha_{in}$ and $\alpha_{out}$, the slope of radial density change indicated by a power law interior and exterior to $R_0$ respectively. The choice of $\alpha_{in}$ and $\alpha_{out}$ (here 3 and -3) doesn't affect our model much in this case because this is a highly inclined disk \citep{lawson2021}.
    \item $\beta$ and $\gamma$ are used to describe the flaring of the disk models and their density distribution. Here we use a disk model with linear flaring, $\beta = 1$ and Gaussian vertical density distribution, $\gamma = 2$.
\end{enumerate}

For a given disk model, the fractional loss of flux due to PSF-subtraction effects is robust to an overall scaling factor. So, rather than varying disk brightness as a parameter, we determine the optimal brightness for each model by computing the error-weighted least-squares normalization between the processed model and the data.

\subsection{Results of Forward Modeling}

\begin{figure*}
    \centering
      \includegraphics[trim=1.5cm 0.5cm 2cm 0.5cm,clip,width=\textwidth]{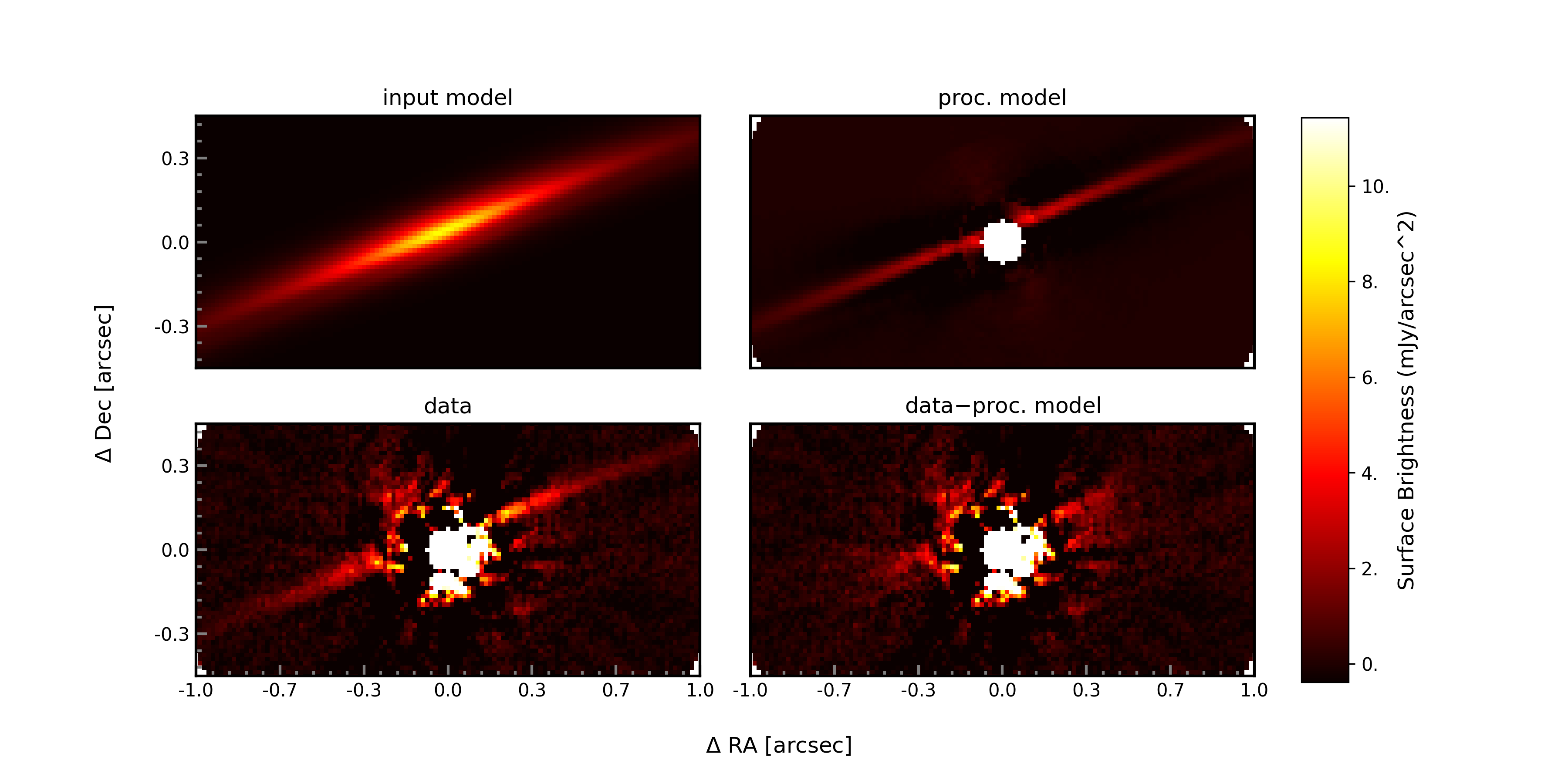}
      \caption{Forward modeling results. From left to right: input model-- the best fitted synthetic disk which is the input model convolved with the instrumental PSF prior to forward modeling; proc model-- the model disk post forward modeling, which yields the synthetic equivalent of the wavelength collapsed image of the disk after ADI A-LOCI PSF subtraction; data-- the wavelength collapsed, PSF subtracted image of the disk; data-model-- the residual for the data product and the synthetic disk model post processing. The minimal structure that remains after subtracting the two disk images suggests our model closely matches the data.}
    \label{fig:fwdmod_panel}
\end{figure*}

\begin{figure*}
    \centering
      \includegraphics[width=\textwidth]{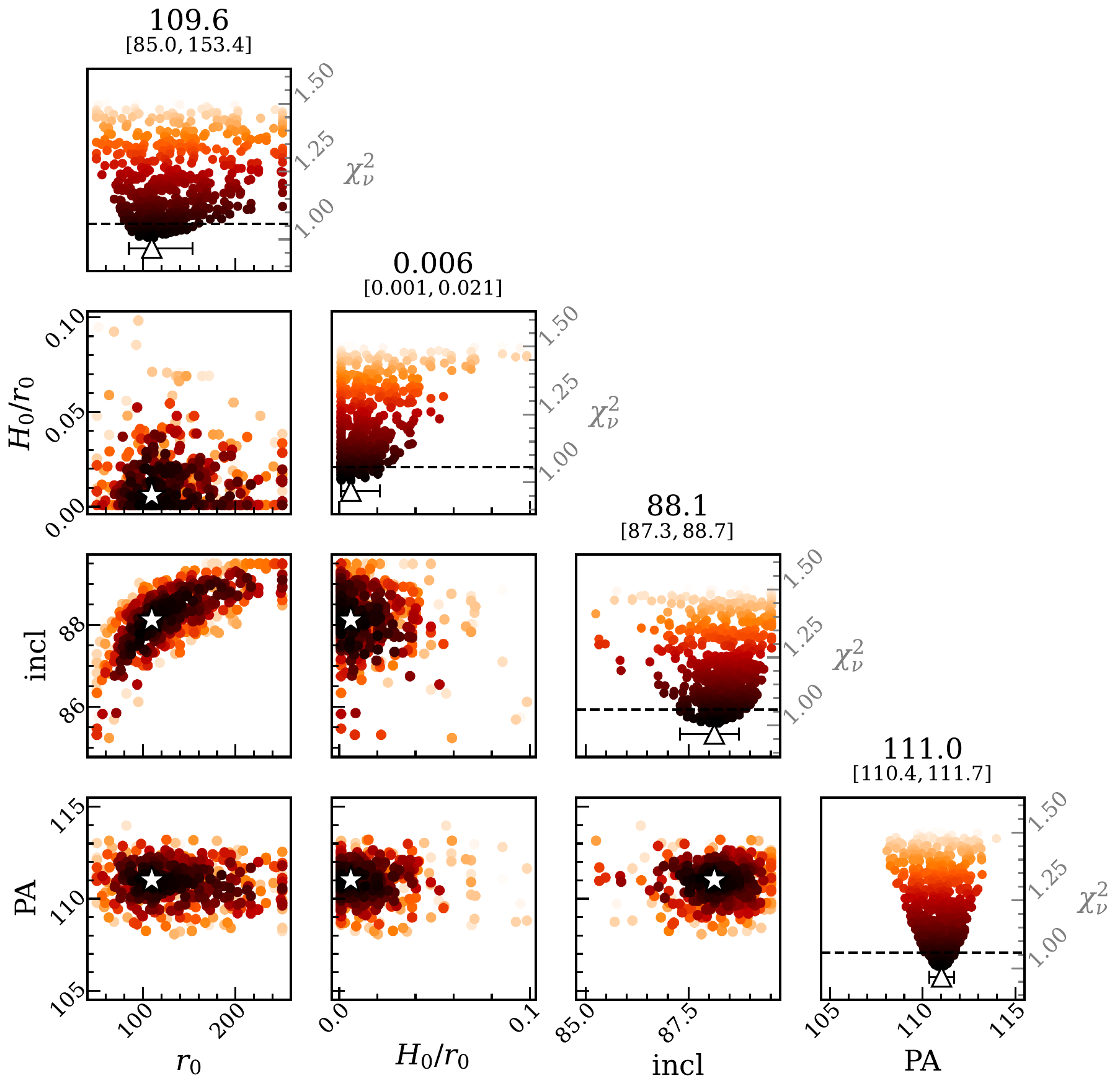}
      \caption{Corner plot showing the optimization of the disk model to the ADI-ALOCI result. The star and triangle symbols show the best model DE converged at in the off-diagonal and diagonal panels respectively. The diagonal panels show the solution as a function of $\chi_\nu^2$. The off-diagonal panels display the solutions as a function of two of the parameters. Each point in the subplots represents a single sample, colored according to its ${\chi }_{\nu }^{2}$ score that ranges from $\chi_{\nu,min}^2$ to $\chi_{\nu,min}^2 + 10 \cdot \sqrt{2/\nu}$. The dotted black line in the diagonal panels mark the threshold for acceptable solutions.}
    \label{fig:cornerplot}
\end{figure*}

Table \ref{tab:fwdmod_vals} gives the values of key parameters DE converged on that describes the structure of the disk after forward modeling the synthetic disk models. In addition to the overall best fit, with $\chi_\nu^2 = \chi_{\nu,min}^2$, we also identify a subset of models which provide acceptable fits to the data based on the prescription from \cite{thalmann2013}:
\begin{equation}
    \chi_\nu^2 \leq \chi_{\nu,min}^2 + \sqrt{2/\nu}
\end{equation}
where $\nu$ is the number of degrees of freedom. The degrees of freedom is defined as the difference between the number of resolution elements in the binned region of interest and the number of free model parameters, and is equal to 808 here. Figure \ref{fig:fwdmod_panel} shows the synthetic disk, followed by how it looks after processing, and compares it to the wavelength collapsed science image of the disk. The synthetic disk is the input model with the parameters from Table \ref{tab:fwdmod_vals} convolved with the instrumental PSF. The rightmost image in Figure \ref{fig:fwdmod_panel} (titled data-model) shows that the model matches the disk observation well, as very little structure is left after subtracting the model from the PSF-subtracted science image, which matches with the noise profile from SNR maps. Figure \ref{fig:cornerplot} shows a corner plot that displays the range of values explored by DE in the parameter space, the solution as a function of $\chi_\nu^2$ and the correlation between the different parameters. 

\begin{table}[t]
    \centering
    \caption{Best-fit parameters}
    \small{
    \begin{tabular}{cc}
      \hline
      \hline
      Parameter & Value \\
      \hline
       $R_0$ (au) & 109.6$^{+43.8}_{-24.6}$ \\
       $H_0/R_0$ & 0.006 \\
       $i$ ($\arcdeg$) & 88.1$^{+0.6}_{-0.8}$ \\
       $PA$ ($\arcdeg$) & 111.0$^{+0.7}_{-0.6}$ \\
       \hline
    \end{tabular}}
    \label{tab:fwdmod_vals}
\end{table}

\section{Results}
\subsection{Disk geometry}

The results from forward modeling indicate that the disk is highly inclined. Table \ref{tab:compare_vals} compares our results with those reported by \citet{hinkley2021} from their 2015 data. As seen in the table, the peak radius $R_0$ is different, whereas the inclination $i$ and position angle $PA$ are nearly the same. The difference in radius is harder to understand since the dust was modeled as a narrow ring in each case. The width of the ring (defined using the the slope of radial density change) is not well constrained for highly inclined disks but the larger value for $R_0$ that we find suggests an extended disk. $R_0$ is larger here (though within the errors) perhaps because the scattered light model of the disk is fitted for J, H, and K bands in this study, whereas \citet{hinkley2021} fits the model for data at 2.2$\mu$m, which corresponds to the K band. There is greater scattering at the shorter wavelengths corresponding to the J and H bands, and we hence see a larger disk. To better constrain the radius, and consequently the ring-structure of the disk, we should fit a model individually for each of J, H and K bands and see how $R_0$ changes in each case and compares with the value quoted by \citet{hinkley2021}. In this work, we fit a model for the K-band and report the results in Table \ref{tab:compare_vals}. The discrepancy in $R_0$ can also be due to the ansae of the disk being comparable to our field of view. The viewing geometry is also not optimal to measure $R_0$ precisely \citep{hinkley2021}. Also, \citet{hinkley2021} uses the Henyey-Greenstein SPF, which is quite different from Hong SPF. This can cause a difference in the final parameters, such as $R_0$ here.

\begin{table}[]
    \centering
    \caption{Comparing disk geometry as derived in this paper in broadband and K-band to values derived by \cite{hinkley2021}}
    \small{
    \begin{tabular}{cccc}
         \hline
         \hline
         Parameter & Broadband & K-band & Hinkley \\
         \hline
         $R_0$ (au) & 109.6$^{+43.8}_{-24.6}$ & 95.0$^{+64.9}_{-24.2}$ & 68.0$^{+20}_{-11}$  \\
         $i$ ($\arcdeg$) & 88.1$^{+0.6}_{-0.8}$ & 87.9$^{+0.7}_{-1.2}$ & 86.1$^{+2.2}_{-2.0}$ \\
         $PA$ ($\arcdeg$) & 111.0$^{+0.7}_{-0.6}$ & 111.0$^{+1.1}_{-0.8}$ & 110.2$\pm$1.2 \\
         \hline
    \end{tabular}} \label{tab:compare_vals}
\end{table}

Since the Keck data was taken at $2.2\,\mu$m, we forward modeled the K-band data as well to see if the results compare better to the NIRC2 data. While the values of $i$ and $PA$ are consistent with the forward-modeling results of the broadband data, we see that $R_0$ is smaller and closer to that reported by \citep{hinkley2021}, though the error is large due to the relatively low signal-to-noise ratio in this narrow band where the scattering is least efficient.

Figure \ref{fig:cornerplot} shows that the fitted radius and inclination are correlated but that the other parameters are independent of each other.
Compared to a more computationally demanding MCMC search, DE rapidly approaches the global minimum which is a computational advantage but without providing rigorous statistical uncertainties for fit parameters. 

\subsection{Limits on Planet Detection}

\begin{figure*}
    \centering
2-00090    \includegraphics[width=\textwidth]{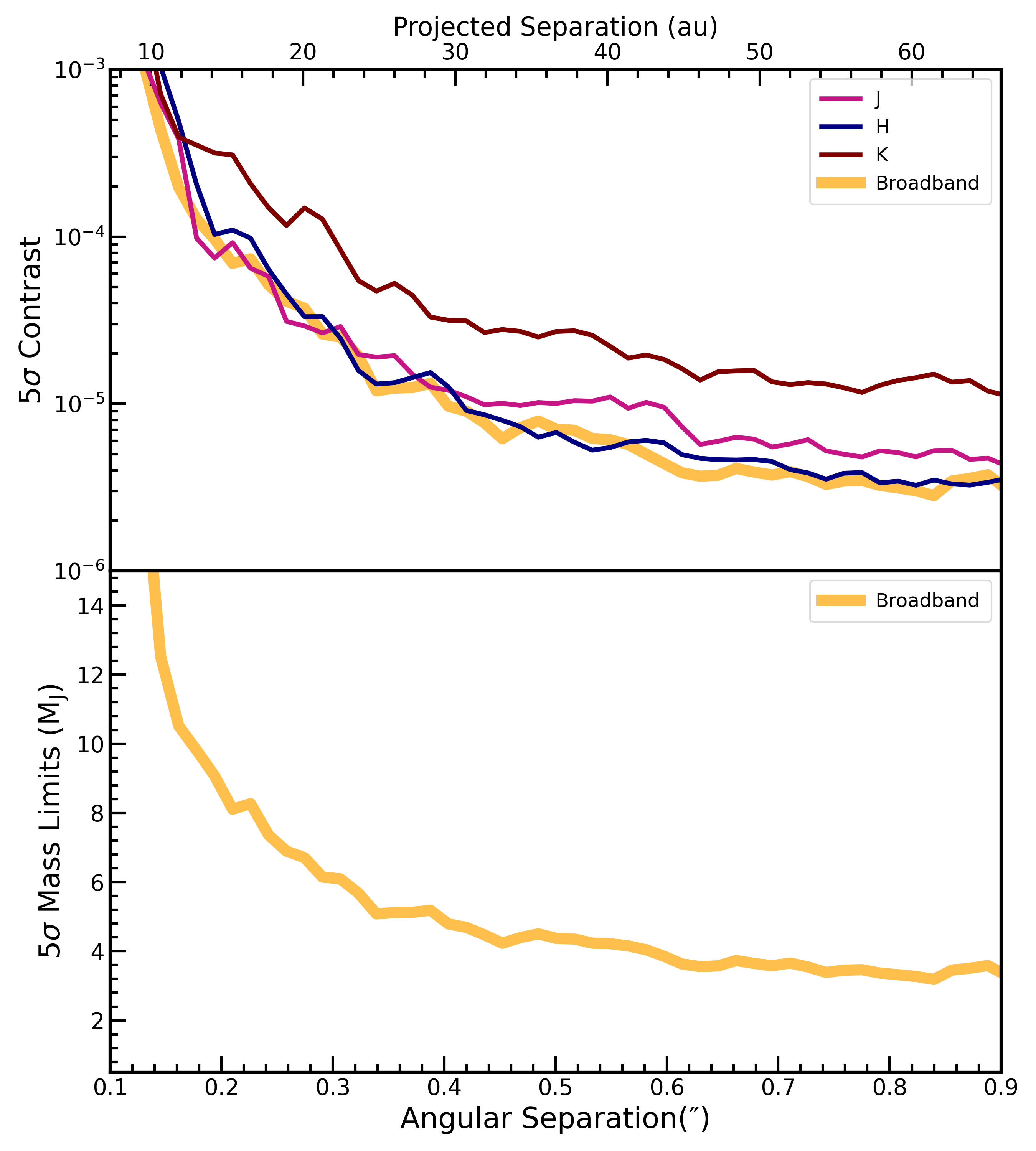}
    \caption{The contrast curve curve for point sources at different wavelengths. The 5$\sigma$ limit on the intensity relative to the central star is plotted as a function of the stellocentric angular separation in arcseconds and stellocentric separation in au.}
    \label{fig:contrast-curve}
\end{figure*}

SCExAO/CHARIS has been used to discover super- Jovian planets and brown dwarfs around nearby stars \citep{kuzuhara2022,steiger2021,swimmer2022,currie2022a,currie2023,tobin2024}. Within our field of view (an annulus covering $\rho$ $\approx$ $0\farcs{}1-1\arcsec$), our data do not reveal any point source signals. To set limits on the presence of substellar companions, we compute 5$\sigma$ contrast limits in J, H, K, and broadband wavelengths as outlined in \cite{currie2018}. For this, we divided the 5$\sigma$ residual noise profile in the wavelength-collapsed ADI image by the median stellar flux. We performed 5 iterations of forward modelling and interpolated the results to create a ``flat-field" (i.e. a map of the synthetic companion throughput vs. position on the detector) which was used to correct the residual noise profile. The contrast curves are shown in Figure \ref{fig:contrast-curve}. No companions were found within contrast levels $\sim 10^{-4}$ at $0\farcs2$ to $< 10^{-5}$ beyond $0\farcs4$.

We then converted the broadband contrast limits to the corresponding masses. We used the \citet{baraffe2003} evolution models to construct a grid of gravities and temperatures for companions with a range of masses, linearly interpolating between grid values at 20 and 30 Myr.  To determine the flux density in broadband (spanning J, H and K filters) at the predicted values of gravity and temperature, we use the BT-Settl cloudy atmosphere models \citep{allard2011,allard2012,allard2014}. Values ranged from $T_{\rm eff}$ $\sim$ 600 K, $\log(g)$ = 3.5 for a 1 $M_{\rm J}$ planet to $T_{\rm eff}$ $\sim$ 2300 K, $\log(g)$= 4.0 for a 15 $M_{\rm J}$ planet. The broadband contrasts translate to companion mass limits of $\sim 10 M_{\rm Jup}$ within 12 au of the star and $\sim 4 M_{\rm Jup}$ beyond 38 au. 

At these scales, we achieved a deeper contrast than \cite{hinkley2021}, who report that their observations are sensitive to $\sim 8 M_{\rm Jup}$ objects at 50 au, and $\sim 3 M_{\rm Jup}$ objects at $\sim$100 au.  Our true gain in sensitivity is slightly larger, since \citeauthor{hinkley2021} adopt cloudless atmosphere models to predict contrast limits. Clouds modeled in BT-Settl depress near-IR flux densities over the age/companion mass range relevant for both our BD+45$^{\arcdeg}$598 data sets \citep[e.g.][]{currie2011}.  Directly-imaged planets and brown dwarfs in the $\beta$ Pic Moving Group with estimated masses relevant for our limits show evidence for clouds \citep[e.g.][]{bonnefoy2013,currie2013,liu2013,rajan2017,mesa2023}.   

It is not necessarily surprising that we did not directly detect a planet given the sensitivity of these observations, considering super- Jovians are relatively rare based on searches for exoplanets through radial velocity, transit, and direct imaging techniques \citep{currie2023b}. Higher contrast imaging is required to detect lower mass companions and higher fidelity imaging to search for indirect signatures through dust asymmetries.

\subsection{Disk Brightness}

\begin{figure*}
    \centering
    \includegraphics[trim=0.8cm 0.5cm 2cm 1.5cm,clip,width=\textwidth]{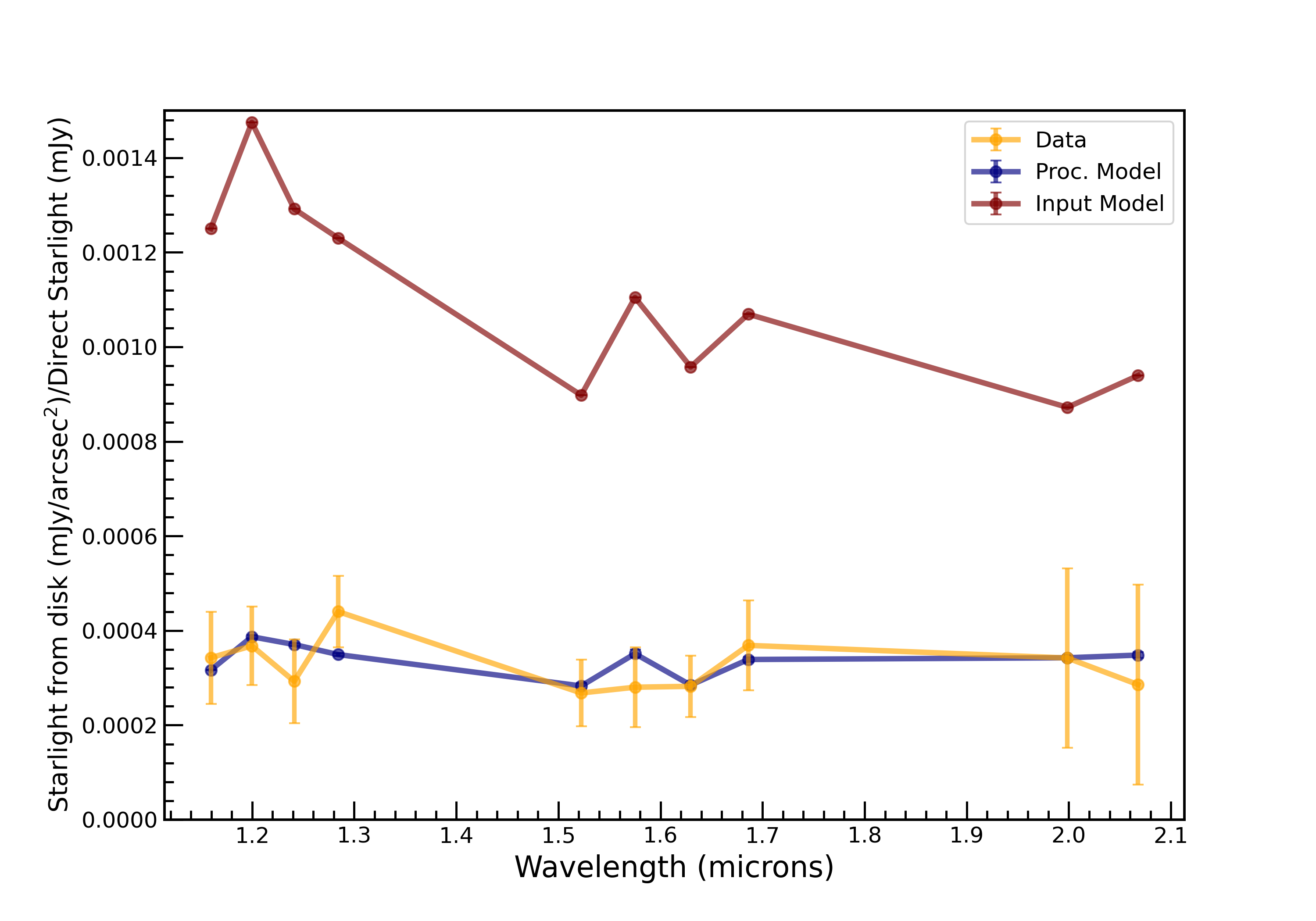}
    \caption{The disk intensity relative to the stellar flux as a function of wavelength for radii $\geq 0\farcs7$. The orange line shows the measured spectrum and the maroon line shows the best fit scattered light model prior to PSF convolution and forward modelling. In the models used here, the disk geometry is kept fixed, while the disk brightness varies as function of wavelength. The blue line shows the model after passing through the same image processing steps as applied to the data.}
    \label{fig:color}
\end{figure*}

The IFU capability of CHARIS provides disk images in 22 independent wavelength bands across the JHK window. Figure~\ref{fig:color} plots the disk spectrum at radial separations $> 0\farcs7$ where there is minimal contamination from image artifacts due to imperfect subtraction of the stellar signal. The intensity was measured for each wavelength channel as a mean of pixel values from $0\farcs7$ to $1\farcs0$ separation from the  host star on either side of the disk. The error bars in the plot represent the standard deviations of the photometry measured in multiple apertures of the same size over the background noise at different angular positions \citep{lawson2020}. The excluded wavelength channels are those affected by telluric absorption or lie outside the nominal bandwidths, and have low SNR values. The spectrum for the processed data is comparable to that for the processed model. But for the input model, the intensity drops with increase in wavelength similar to that reported by \cite{esposito2020} in a large survey of debris disks using Gemini/GPI, indicating blue J-K color and H-K colors, as opposed to the red H-K color for the processed model and the data. 

\begin{figure*}
    \centering
      \includegraphics[trim=0.2cm 1.4cm 1.7cm 1.5cm,clip,width=\textwidth]{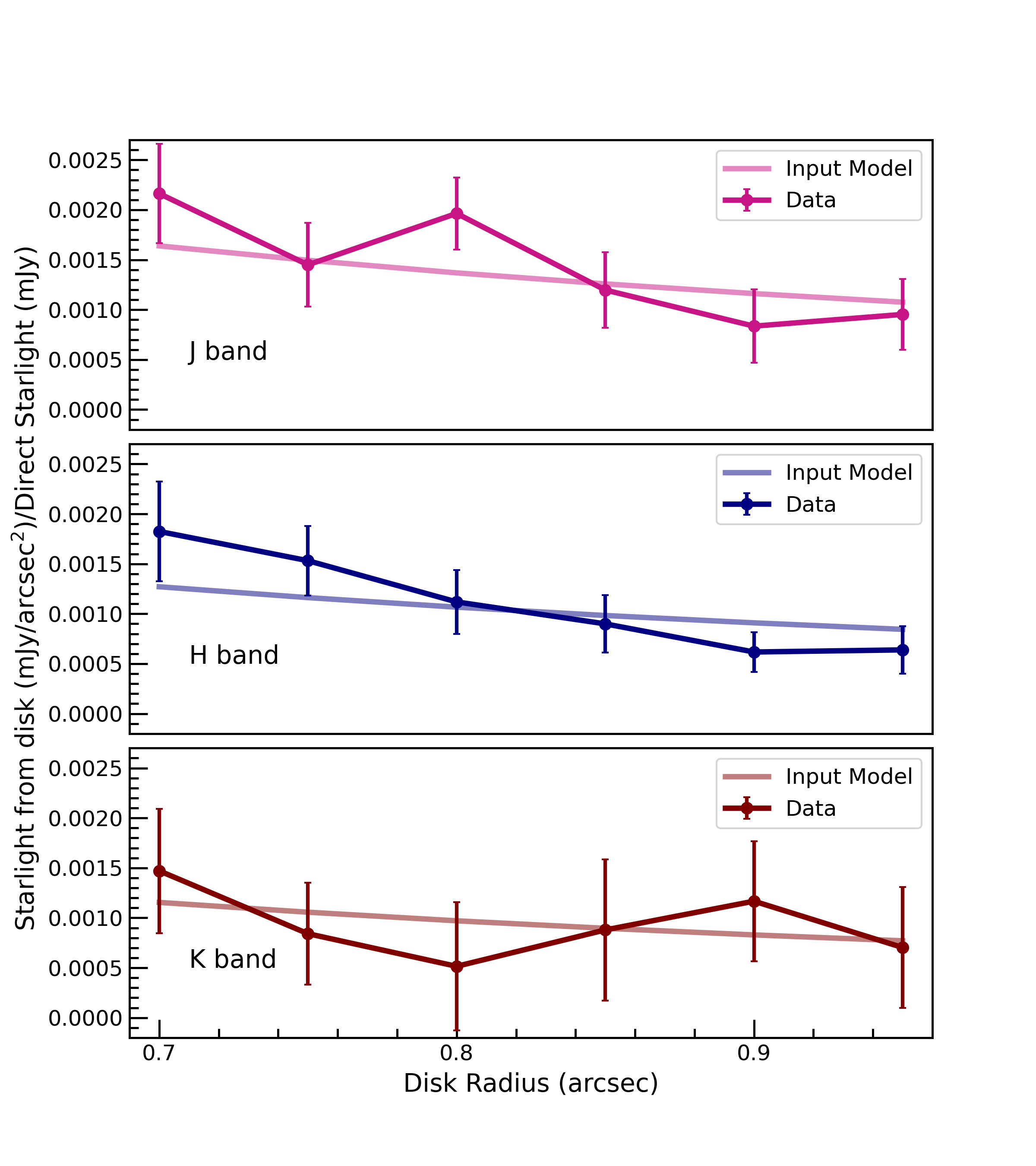}
    \caption{The disk intensity relative to the stellar flux as a function of wavelength for radii $0\farcs7$ to $1\farcs0$. In the models used here, the disk geometry is kept fixed, while the disk brightness varies as function of radius. The disk brightness from the data plotted here has been corrected for self subtraction, and is measured in apertures of width $0\farcs05$, starting from $0\farcs7$ as indicated by the x-coordinate of the data points, and height $0\farcs05$.}.
    \label{fig:color2}
\end{figure*}

The color discrepancy is attributed to self-subtraction effects. This is mitigated in the data by multiplying the disk brightness with a ratio of the disk brightness measured in the input model to that in the processed model. Figure \ref{fig:color2} plots the disk brightness relative to the stellar flux as a function of radial separation from the host star in J, H, and K bands. The intensity was measured for the wavelength channel within each as a mean of pixel values within each of the 5 contiguous $0\farcs05 \times 0\farcs05$ rectangular regions along the disk spine from $0\farcs7$ to $1\farcs0$. The error bars in the plot, as described previously, represent the standard deviations of the photometry measured in mutliple apertures of the same size over the background noise at the same separations from the host star, but different angular positions, as the apertures measuring the disk brightness. 

\begin{figure*}
    \centering
      \includegraphics[trim=0.7cm 0.4cm 1.7cm 1.2cm,clip,width=\textwidth]{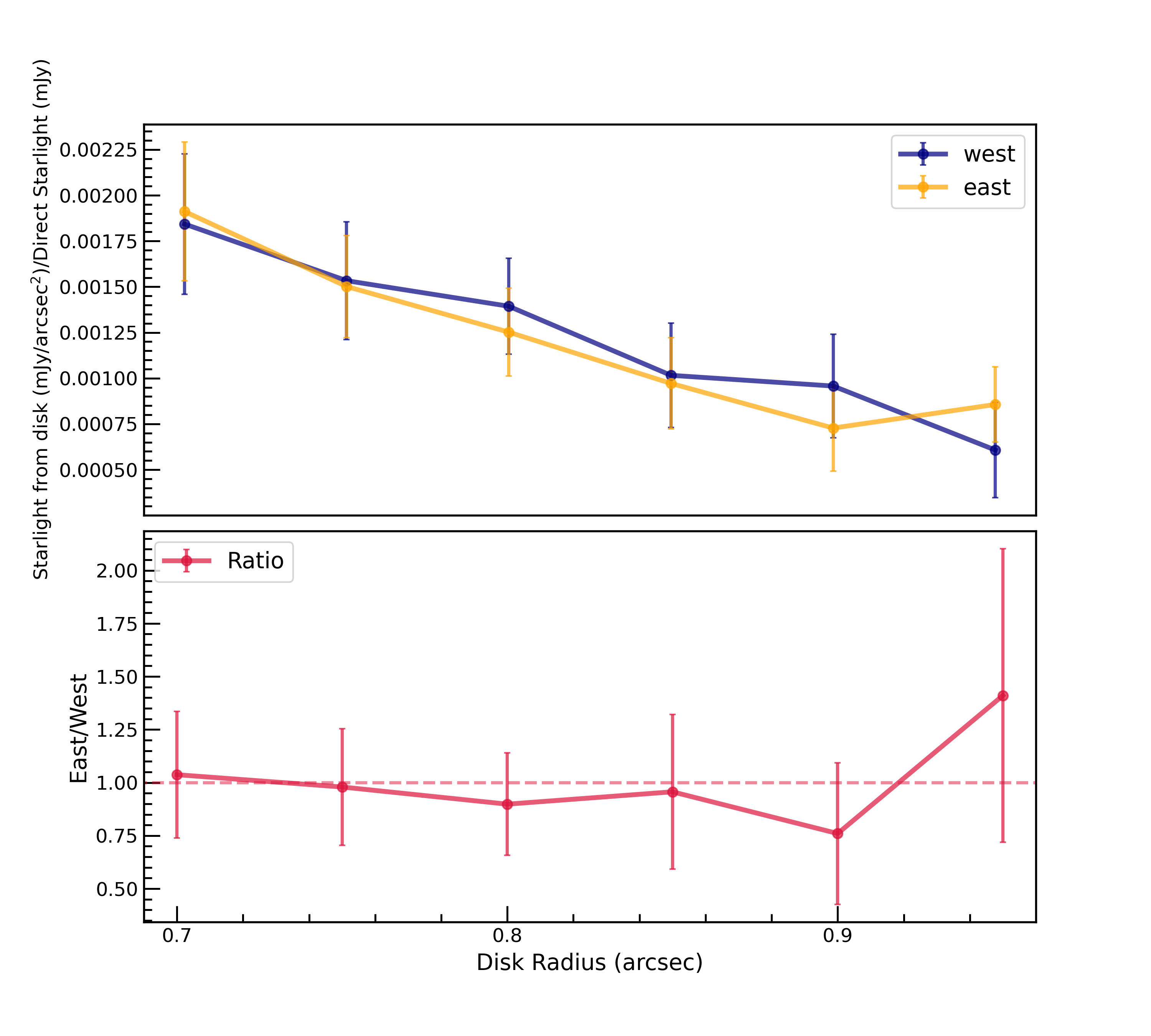}
    \caption{Top: Comparison of surface brightness the east and west sides of the disk spine. The east side is brighter than the west. Bottom: Ratio of the east and west sides, which ascertains that there is no notable brightness asymmetry in the disk.}
    \label{fig:asymmetry}
\end{figure*}

The disk brightness has been calculated separately for the east and west parts of the disk spine to ascertain the asymmetry in the disk structure. Figure \ref{fig:asymmetry} compares the brightness on the eastern and western sides of the disks for a radial separations from $0\farcs7$ to $1\farcs0$. There is no notable asymmetry, as the ratio of the brightness of both sides of disk are equal to or close to 1. We chose not include measurements within $0\farcs7$, as these are affected by residual starlight.

\begin{figure*}
    \centering
    \includegraphics[width=0.8\textwidth]{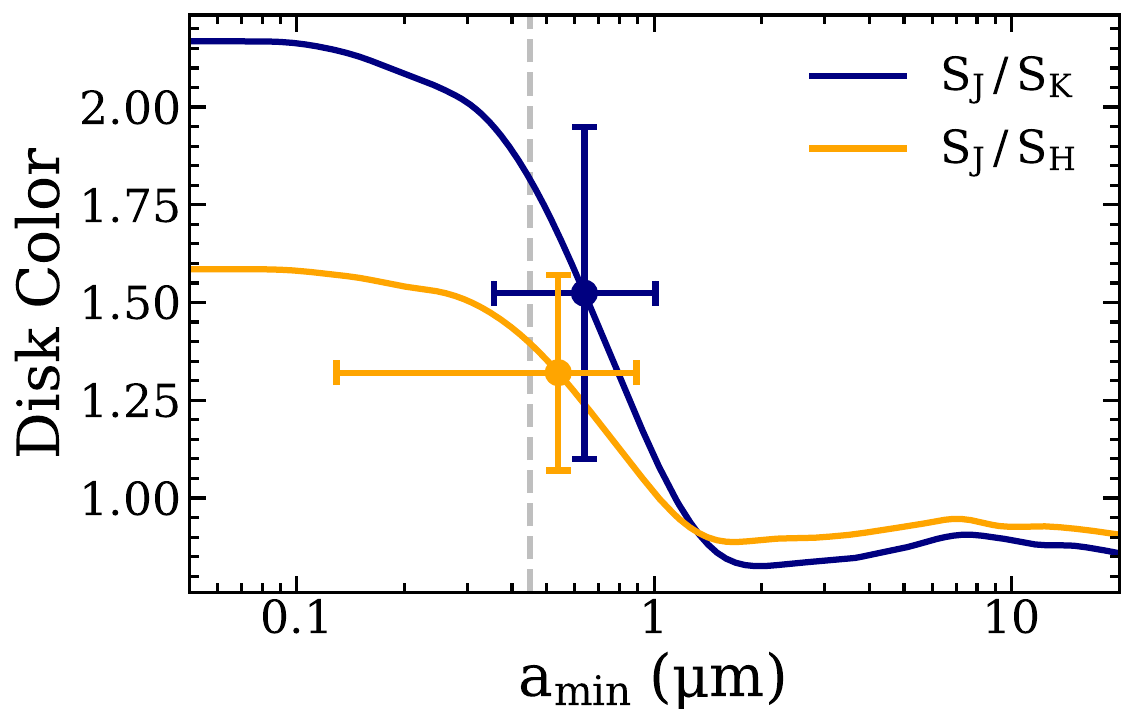}
    \caption{Range of values for minimum grain size assuming a grain composition of standard astronomical silicates derived using agglomerated debris particle (ADP) calculations from \cite{arnold2022}. The solid lines correspond to the variation of disk color as a function of minimum grain size derived from the input models, and the markers (with error bars) are the minimum grain size values derived from the disk color measured in the data corrected for self-subtraction. The grey line marks the blowout size for the star which is approximately $0.45 \micron$.}
    \label{fig:grain_size}
\end{figure*}

To estimate the minimum size of dust grains within the disk, we assume a composition of standard astronomical silicates and adopt the agglomerated debris particle (ADP) data from calculations in \citet{arnold2022}. Assuming a size distribution $n(a) \propto a^{-3.5}$ with a maximum grain size of 200$~\micron{}$, we explore minimum grain sizes ($\rm a_{min}$) in a logarithmically-spaced grid spanning 0.05–20$~\micron{}$ — computing the corresponding disk colors in each case. Figure \ref{fig:grain_size} shows these results, with measurements of the disk from this work overlaid. The best fits to the $S_{J}/S_{K}$ (ratio of brightness in approximate J and K bands) and $S_{J}/S_{H}$ (ratio of brightness in approximate J and H bands) occur for an $\rm a_{min}$ range of $0.36~\micron{} - 1.01~\micron{}$ and $0.13~\micron{} - 0.89~\micron{}$ respectively. The blowout size for the host star, which is $\sim  0.45 \mu$m, falls within the range of $\rm a_{min}$ values. A minimum grain size below the blowout size suggests alterations in grain collision physics when approaching the threshold of small grains \citep{hughes2018}. It may also indicate replenishment of small grains through planetesimal collisions \citep{hahn2010, lawson2020}. A minimum grain size derived from observations, when larger than the blowout size, is likely a result of a change in microphysics of collisional dust production \citep{hughes2018,pawellek2015}.

\section{Conclusions and Future Work}

We have studied  multi-wavelength imaging of the debris disk around BD+45$\arcdeg$598, a nearby, young K1 type star in the $\beta$ Pic moving group. We detect the disk across the J, H, and K bands and fit a scattered light model using a forward modeling algorithm. The inferred geometry of an inclination $i = 88.1\arcdeg$ and position angle $PA = 111.0\arcdeg$ agrees with a previous study by \cite{hinkley2021} using Keck at K-band, though we find a larger radius $R_0 = 109.6$ au likely because our observations extend to shorter wavelengths where the scattering is more efficient and we use a different scattering phase function.

There is no direct evidence for a substellar companion in the disk. However, the 5$\sigma$ contrast limits computed within the CHARIS field-of-view show that there can be $\sim 10 M_{\rm Jup}$ within 12 au of the star and $\sim 4 M_{\rm Jup}$ beyond 38 au. Finally, there is no notable asymmetry in the disk brightness across wavelengths. disk is brighter at shorter wavelengths. The disk color constrains the minimum dust grain size within a range of $\sim$0.13 to 1.01 $\mu$m, which includes blowout size of the star, 0.45 $\mu$m.

In the future, we can use  polarized intensity data to see if this can help improve the disk model. In regard to future observations, SCExAO now has a new NIR pyramid wavefront sensor which allows even higher contrast imaging of fainter targets \citep{lozi2022}. This opens up new areas of investigation for disk imaging and searches for protoplanets. 

This research is based on data collected at the Subaru Telescope, which is operated by the National Astronomical Observatories of Japan. We are grateful for the opportunity to conduct observations from Maunakea, a site of great cultural, historical, and natural significance for the indigenous Hawaiian community. We wish to acknowledge the indispensable role of current and recent Subaru telescope operators, day crew, computer support, and office staff employees in continued successful operation of Subaru. This work has used  NASA’s Astrophysics Data System Bibliographic Services. This research made use of Astropy, a community-developed core Python package for Astronomy \citep{astropycollab2013,astropycollab2018}. 

\bibliography{bibliography}{}
\bibliographystyle{aasjournal}

\newpage

\section*{Figures}

\newpage

\newpage

\newpage

\newpage

\newpage

\newpage

\begin{figure*}
    \centering
    \includegraphics[width=\textwidth]{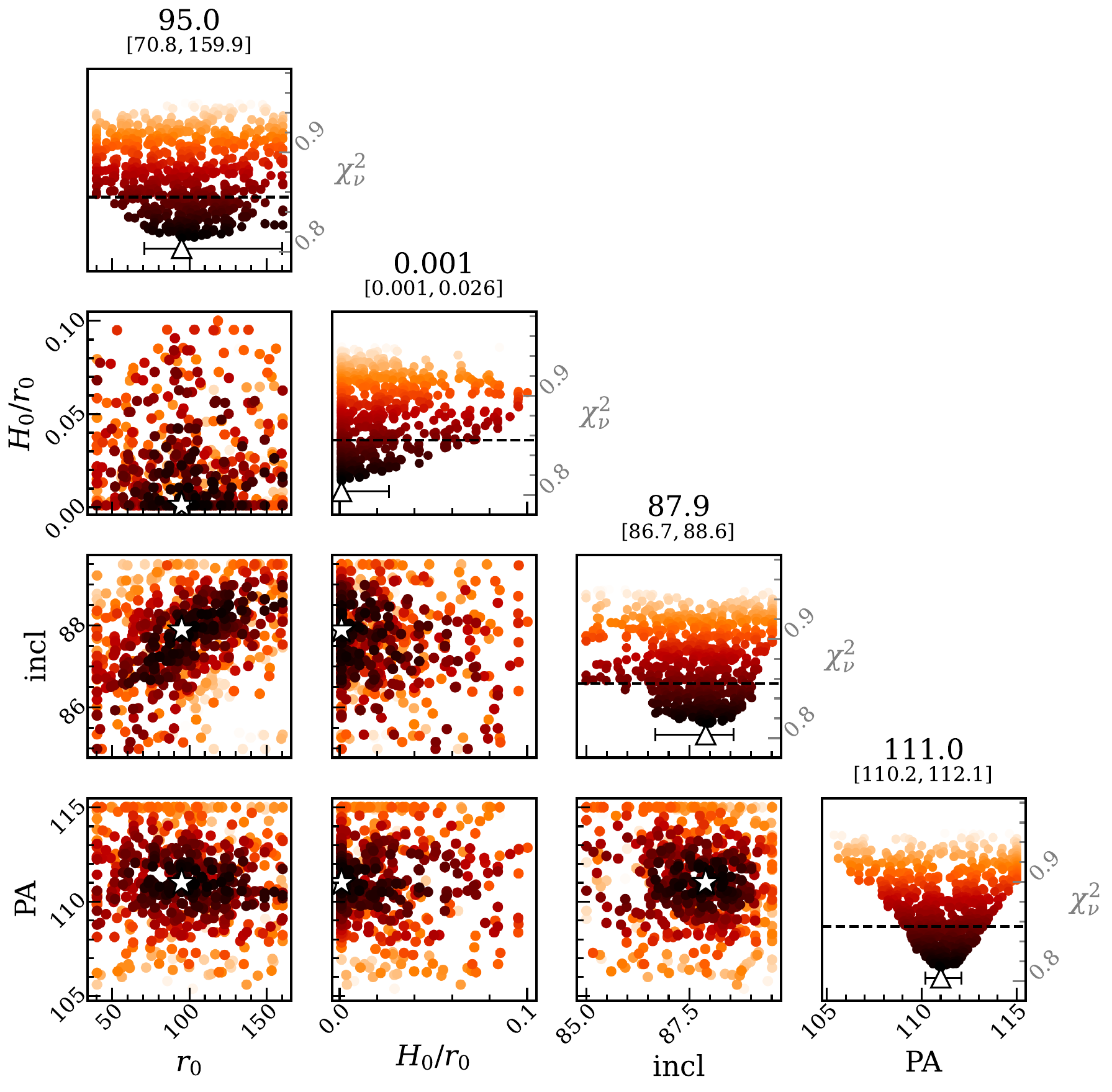}
      \caption{Corner plot showing the optimization of the disk model to the ADI-ALOCI result of the slices of the image cube corresponding to $\lambda$ = 1.93, 2.0, and 2.07 microns.}

    \label{fig:cornerplot_k}
\end{figure*}

\end{document}